\documentclass[%
reprint,
nofootinbib,
 amsmath,amssymb,
 aps,
]%{article}
{revtex4-1}
\usepackage{epsfig,epsf,rotating,wrapfig,tabularx}
\usepackage{dcolumn}% Align table columns on decimal point
\usepackage{bm}% bold math
\usepackage{graphicx}% Include figure files
\begin{document}

\preprint{APS/123-QED}

\title{Transverse momentum distributions of baryons at LHC energies}% Force line breaks with \\

\author{\firstname{A.~A.}~\surname{Bylinkin}}
 \email{alexandr.bylinkin@cern.ch}
\affiliation{%
Moscow Institute of Physics and Technology, MIPT, Moscow, Russia\\
National Research Nuclear University MEPhI, Moscow, Russia
}%
\author{\firstname{O.~I.}~\surname{Piskounova}}
 \email{piskoun@lebedev.ru}
\affiliation{%
P.N.Lebedev  Physical Institute, LPI, Moscow, Russia
}%


\begin{abstract}
Transverse momentum spectra of protons and anti-protons from RHIC ($\sqrt{s}$ = 62 and 200 GeV) and LHC experiments ($\sqrt{s}$= 0.9 and 7 TeV) have been considered. The data are fitted in the low $p_T$ region with the universal formula that includes the value of exponent slope as a main parameter. It is seen that the slope of low-$p_T$ distributions is changing with energy. This effect impacts on the energy dependence of average transverse momenta, which behaves approximately as $s^{0.06}$ that is similar to the previously observed behavior for $\Lambda^0$-baryon spectra. In addition, the available data on $\Lambda_c$ production from LHCb at $\sqrt{s}= 7$ TeV were also studied. The estimated average $<p_T>$ is bigger than this value for protons proportionally to masses. The preliminary dependence of hadron average transverse momenta on their masses at LHC energy is presented.
\end{abstract}

\pacs{Valid PACS appear here}% PACS, the Physics and Astronomy
\maketitle

%\tableofcontents
\section{Introduction}
The transverse momentum distributions are the primary data that can be obtained in the study of hadron spectra at the modern collider experiments. Interpretation of these distributions as in strongly theoretical perturbative QCD approaches as in various phenomenological models can shed a light upon the physics of hadroproduction processes at high energies. We apply the phenomenological approach based on the previous attempts of the description of $p_T$ spectrum in the framework of Quark-Gluon String Model \cite{QGSM}. 

The model has described the data of previous colliders up to energies $\sqrt{s} = 200$ GeV at the area of low $p_T$'s that gives main contribution to the average value of transverse momenta. Recently $\Lambda^0$-hyperon production has been studied \cite{Lambda} in terms of the QGSM. We suggest here to compare the latest data measured at LHC on proton production and to show the resulting average transverse momenta, $<p_T>$, as a function of c.m.s. energy, $\sqrt{s}$, of colliding protons.
It seems interesting as well to compile the available data on $<p_T>$ at LHC energy for all sorts of hadrons and to present them as the function of produced hadron mass.

\section{Models for the proton spectra}
Let us first describe the QGSM approach, which has been applied for recent studies of $\Lambda^0$. According to this approach the spectra of baryon production can be parameterized in the following way:
\begin{equation}
\label{qgsm}
\frac{d\sigma}{p_T d p_T} = A_0\exp [-B_0\cdot (m_T - m_0)],
\end{equation}
where, $m_T=\sqrt{p_T^2+m_0^2}$, $m_0$ is the mass of the produced hadron and $B_0$ is the slope parameter for the considered energy. In the early paper \cite{Lambda}, it was also shown that the value of the slope parameter $B_0$ becomes dependent on the collision energy.

As the next approach, the widely known Tsallis parameterization~\cite{Tsallis1, Tsallis2} that is giving rather good description of hadroproduction spectra might be used. Transverse momentum distribution can be expressed in this model by the following formula:
\begin{center}
\begin{equation}
\label{tsallis}
\frac{d\sigma}{p_T dp_T} = \frac{A}{(1+\frac{E_{Tkin}}{TN})^N}
\end{equation}
\end{center}
with a temperature-like parameter $T$ and the power-like $N$. In this forumula~(\ref{tsallis}) $E_{Tkin}$ - is the transverse energy that can be calculated in the following way: 
\begin{center}
\begin{equation}
E_{Tkin} = \sqrt{p^2_T + M^2} - M,
\end{equation}
\end{center}
with $M$ equal to the hadron mass.

Recently a two component model for hadroproduction has also been developed~\cite{OUR1}. It was suggested to parameterize the large variety of charged particle spectra by a sum of an exponential (Boltzmann-like) and a power-law $p_T$ distributions:
\begin{equation}
\label{eq:exppl}
\frac{d\sigma}{p_T d p_T} = A_e\exp {(-E_{Tkin}/T_1)} +
\frac{A}{(1+\frac{p_T^2}{T_2^{2}\cdot n})^n},
\end{equation}
In~\cite{OUR2,OUR3} it have been concluded, by the way, that spectra of baryon production can be described at high energies with only power-law term of the equation~(\ref{eq:exppl}).

\begin{figure}
\includegraphics[width=8cm]{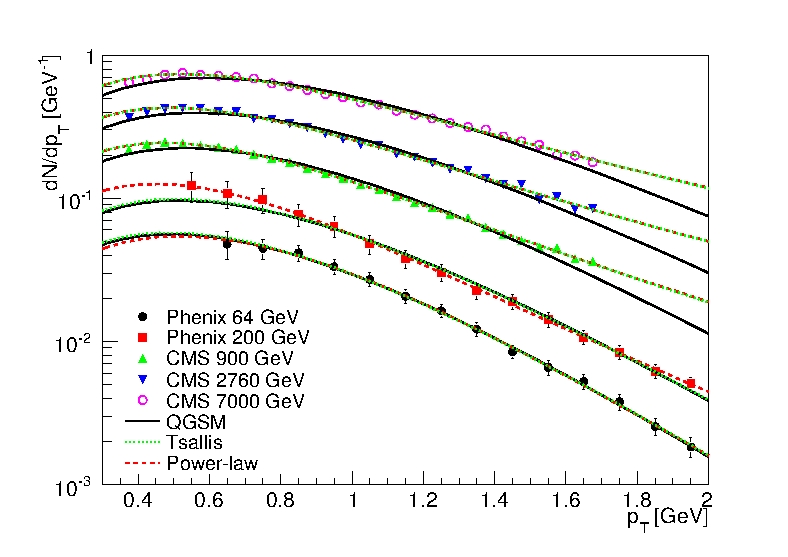}\\
\caption {\label{fit}} Proton production spectra~\cite{PHENIX, CMS} shown with arbitrary normalization together with various phenomenological approaches. Black solid line shows the QGSM~(\ref{qgsm}), green pointed line - Tsallis-fit~(\ref{tsallis}) and red dashed line - the power-law~(\ref{eq:exppl}).
\end{figure}

The results of these three approaches are shown in the figure~\ref{fit} in the comparison with the available experimental data on proton production in $pp$ collisions at different energies. 

It should be noticed after all, that these three approaches have the exponential-like behavior in the low-$p_T$ region and give a reasonable description of the experimental data. Therefore, for the analysis suggested in the present paper considering the mean transverse momentum $<p_T>$ of produced baryons one can use any of these models, since high transverse momenta particles (not measured by PHENIX and CMS) do not add a big contribution to the mean value.

If we take the experimental data, it is clearly seen in the shapes of transverse momentum dependences that the spectra of produced baryons become harder with the higher energies. Thus, it is interesting to study the variation of the mean transverse momentum $<p_T>$ with the collision energy in order to estimate the power of its growing.

\section{Mean transverse momenta}
\begin{figure}
\includegraphics[width=8cm]{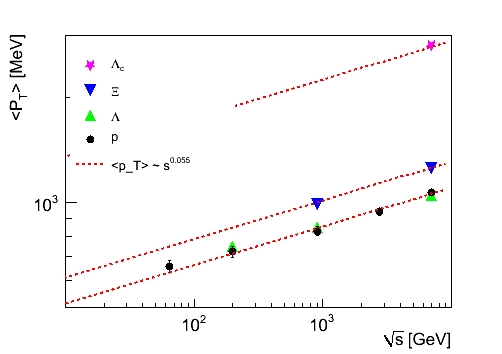}\\
\caption {\label{pt}} Mean transverse momenta of charged baryons~\cite{PHENIX, CMS, CMSL, LHCb} as a function of c.m.s energy $\sqrt{s}$. Lines show the power-law dependece $s^{0.055}$).
\end{figure}

Let us now discuss the mean transverse momenta of produced baryons and look at its dependence on the collision energy, $\sqrt{s}$. It seems reasonable also to compare the values calculated for the proton spectra with other available data on baryon production: $\Lambda$, $\Xi$~\cite{CMSL} and $\Lambda_c$~\cite{LHCb} spectra.
Figure~\ref{pt} shows such dependence for the various experimental data. The steep rise of the mean transverse momenta $<p_T>$ is seen as a function of energy $\sqrt{s}$. Remarkably, this rise can be parameterized by the same power-like $s^{0.055}$ behavior in case of all the species of produced baryons. These observations mean that the transverse distributions of baryons might have the same nature and their variations are not dependent on the quark-composition of the produced baryon.

%Remarkably, in ~\cite{OURM} it was shown that the dependence of the $<p_T>$ extracted for the power-law term of (\ref{eq:exppl}) alone observes practically the same energy dependence??? $s^{0.055}$ as in present analysis.
Remarkably, in \cite{OURM} it was shown that the $<p_T>$, which were estimated with the power-law spectra alone, behaves as $s^{0.35}$.
This dependence would refer to the intercept of supercritical Pomeron, $\delta_P(0)$, that causes the hadroproduction cross section dependence like $s^{\delta_P(0)}~\approx s^{0.12}$. Unfortunately, the growing of $<p_T>$’s have nothing to do with the growth of cross sections, because, first of all, mean transverse momenta are not of the same dimension as cross sections, see the formula \ref{eq:avrpt}. There is no room for pomeron-intercept dependence in the mean transverse momenta definition. 

\begin{equation}
\label{eq:avrpt}
<p_T>=\frac{\int{p_T \cdot \frac{d\sigma}{dp_T}}}{\int{\frac{d\sigma}{dp_T}}}
\end{equation}
The similar misleading appears if we postulate the exponential behavior of spectra at low $p_T$ as “hadron evaporation”, see \cite{kharzeev}. Analyzing the transverse momentum distributions in \cite{antip-p}, it was concluded that spectra in proton-proton and antiproton-proton collisions have different forms in this very region of low-$p_T$. Why antiprotons should “evaporate” so differently than protons? 
The reasons that discussed above make this research more interesting from the point of view of further higher energy hadron experiments.

\begin{figure}
\includegraphics[width=8cm]{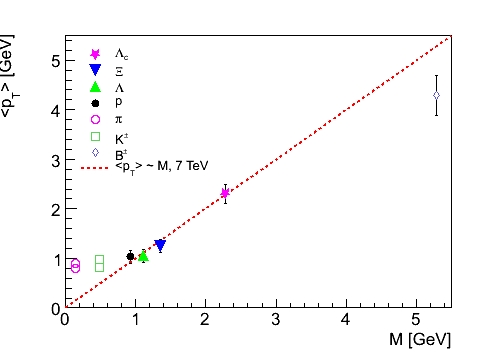}
\caption {\label{m}} Mean transverse momenta of charged baryons~\cite{CMS, CMSL, LHCb, Bmes} as a function of their mass at the energy $\sqrt{s} = 7000$ TeV. Red dashed line shows linear dependence.
\end{figure}

Another interesting implication can be revealed from the comparison of the mean transverse momenta of various produced baryons at the certain collision energy as a function of their masses, shown in figure~\ref{m}.
Let us note that a linear dependence between the mean transverse momenta $<p_T>$ and the baryon mass $M$ is observed. Remarkably, at $\sqrt{s} = 7000$ TeV the transverse momentum reaches the value of baryon mass, t.e. $<p_T> \sim M$. Further measurements at LHC Run-II can shed light on whether or not the average transverse momentum expansion in the baryon production has a limit $<p_T> = M$.
It also might be interesting to compare $<p_T>$’s as a function of produced baryon mass $M$ with the mass dependence of the mean transverse momenta that have been calculated from the description of charged meson production.

\section{Conclusion} 
Transverse momentum spectra of protons and anti-protons from RHIC ($\sqrt{s}$ = 62 and 200 GeV) and LHC experiments ($\sqrt{s}$= 0.9 and 7 TeV) have been described in the QGSM approach. This model seems working for the up-to-date collider energies, because spectra at low $p_T$ are giving the main part of integral cross section. 

The enhancement of power-low contribution into the spectra at high$-p_T~'$s causes the change of low $p_T$ exponential slopes, so that $<p_T>$ are growing with energy.
If we are analyzing the spectra of different baryons their average transverse momenta grow linearly with masses. 
The LHC$_b$  experiment has obtained more data on heavy quark baryons and mesons which are going to supply our analysis.

Nevertheless, the observed changes in hadroproduction spectra seem not to be so dramatic to cause the "knee" in  proton     spectra measured in cosmic rays due to just the interactions into the detector at the laboratory system energies that correspond to the energy range of LHC experiments.

\section{Aknowledgements}
Authors express their gratitude to Prof. K.G. Boreskov and to Prof. M.G. Ryskin for numerous discussions and useful advises.

%% References

\nocite{*}
\bibliographystyle{elsarticle-num}
%\bibliography{martin}



\end{document}